# Wearable Conformal Fiber Sensor for High Fidelity Physiological Measurements

Alessio Stefani, Ivan D. Rukhlenko, Antoine F. J. Runge, Maryanne C. J. Large, and Simon C. Fleming

*Abstract*—Wearable devices are becoming increasingly important, addressing needs in both the fitness and the medical markets. In this paper, we describe a novel sensing platform based on a hollow-core polyurethane optical fiber, operating through capillary guidance, that acts as a conformal sensor of pressure or deformation. The novelty is achieved by combining a simple structure (hollow capillary) and a simple detection technique (intensity-based measurement) with unconventional material properties (extreme deformability and high optical absorbance). Used on the wrist and ankle, the sensor allows detailed features of the cardiac pulse wave to be identified with high fidelity, while on the chest it allows the simultaneous measurement of breathing rate and walking cadence. Used together, an array of such sensors (with others) could be incorporated into clothing and provide physiologically rich real-time data for health monitoring.

## I. INTRODUCTION

WEARABLE medical devices use non-invasive sensors to measure physiological characteristics, including heart rate, oxygen saturation, body temperature, and motion analysis. These devices range from consumer devices such as smart watches, to devices for optimizing sporting performance, to medical devices such as continuous glucose monitoring, fall detection devices, and sleep trackers. Wearables for health monitoring have developed two distinct, though complementary uses: in fitness and wellbeing, and for the management of chronic diseases.

Elite athletes now routinely use wearable sensors to monitor their physiological performance (e.g., devices from Catapult or Zephyr). Millions of consumers use health tracking on devices such as FitBit or Apple Watches. Indeed, the Apple Watch Series 6 was launched with the slogan "the future of health is on your wrist".

More specialized medical wearables have allowed patients with chronic diseases to better manage their conditions. They have also enabled some powerful emerging trends, such as remote patient monitoring, improved home-based care and telehealth — all of which have accelerated during the COVID-19 pandemic.

This growth will require the development of increasingly sophisticated sensor technology to improve device performance. These improvements will encompass several aspects, including miniaturization, cost, ease of integration with other products, and the quality and type of physiological measurements. Sensors that are integrated into clothing, for example, would be less intrusive and more comfortable to use than standalone devices. In other cases, the challenge is to produce hospital-quality measurements that are reliable across a range of body types and physical activities. As an example, one issue that has been raised for optical sensors that rely on light transmission through the skin is that different results are obtained depending on the skin color [1].

Optical fiber sensors are particularly well suited to wearable applications more generally [2]. They can be used to measure a wide range of physical parameters such as temperature [3], mechanical strain [4], and pressure [5]. They can be made compact and lightweight, have a large bandwidth, and can operate in a wide range of environments [6]. These properties make optical fiber sensor-based devices highly suitable for biomedical applications [7, 8]. Importantly, they can also be incorporated into fabrics.

Indeed, fiber devices have become increasingly sophisticated, including electrical, optical, and mechanical components. These now include fibers that incorporate diodes, microelectromechanical systems (MEMS), memory elements and energy storage [9, 10]. The increasing complexity of these devices has been described as a "Moore's Law for Fibers" [9].

However, most of these fiber devices and systems are fabricated from materials (such as glass and stiff plastics) that are much more rigid than most biological tissues. This limits their potential for some physiological measurements, such as those based on conformal contact with the skin or measuring small forces. Materials with lower Young's modulus allow for a greater response to external perturbations and recently a growing number of flexible fibers [11-24] have been demonstrated for implementation in robotics [11, 12] and wearables [13-16].

To achieve the low Young's modulus, various materials have been used, e.g., PDMS [16], hydrogels [17], and various elastomers [11, 12, 18-21] and diverse fabrication techniques, i.e., molding [11, 12, 16, 17, 20], extrusion [21], spinning

---

This work was supported in part by Marie Sklodowska-Curie grant of the European Union's Horizon 2020 research and innovation programme (708860), Australian Research Council (ARC) Discovery Project DP170103537, and the Westmead Applied Research Centre blood pressure challenge. Alessio Stefani and Ivan D. Rukhlenko contributed equally to this work.

Alessio Stefani, Ivan D. Rukhlenko, Antoine F. J. Runge, Maryanne C. J. Large, and Simon C. Fleming are with Institute of Photonics and Optical Sciences (IPOS), School of Physics, The University of Sydney, NSW, 2006, Australia (e-mail: alessio.stefani@sydney.edu.au).

Color versions of one or more of the figures in this article are available online at http://ieeexplore.ieee.org

polymerization [15], and thermal drawing [18, 19]. The combination of materials and fabrication techniques determined the specific mechanical properties (e.g., Young's modulus and elastic limit) as well as the maximum achievable fiber length. What all previous reports have in common is the use of solid waveguides where the light propagates in the material itself and the achievable deformation is fully controlled by the material properties.

We recently reported on the fabrication, by fiber drawing, and characterization of hollow-core polyurethane (PU) — a thermoplastic elastomer — optical fibers, which operate by capillary guidance, relying on glancing incidence reflection [22-24]. Waveguides that confine light into an air core to achieve lower transmission loss date back sixty years [25]. Historically, the complexity of the structures has increased from capillaries to band-gap guiding fibers [26]. However, there has been a recent trend moving towards simpler structures, mostly driven by the advancements on antiresonant fibers [27], all the way back to hollow capillaries [28]. While relatively high loss, such fibers can operate effectively over short distances on the scale of the body and are well suited to integration into clothing. The remarkable material properties of PU allowed high levels of elongation and large deformation (compression and/or bending) that resulted in significant transmission losses [22]. Thus, the sensitive detection of pressure variation or deformation was possible through a simple optical intensity measurement.

The choice of a fiber structure containing a significant air fraction provides a further means to tune the mechanical properties and effectively increase the sensitivity to deformation and in particular to transverse applied pressure. It also adds a degree of freedom to the control of the optical properties.

Hence, the significantly enhanced sensitivity over a greater range of deformations and relative inexpensiveness of hollow-core PU fibers are their key advantages over commercial silica and solid polymer fibers. When compared with electronic transducers and standard capacitive sensors, such fibers offer an additional benefit of noise reduction when measuring pressure due to the smaller area of contact with the measurement site. The significant breaking strain (of up to 600% for thermoplastic PU, TPU) also makes polymer fibers much more rugged than electronics in wearable applications. In this paper, we report two examples of the use of these highly flexible hollow-core fibers as respiration and pulse measurement sensors. In both cases the sensing relies on intensity variation of the transmitted light caused by fiber deformation. We show that breathing can be measured with the integration of a fiber at the chest and that the cardiac pulse wave can be measured accurately with a sensor on the wrist and on the ankle. We believe that the high level of fidelity provided by these novel fibers and their inexpensiveness are the key advantages that will ensure their widespread use in next-generation wearable devices and enable them to replace standard optical fiber sensors in the field of heart rate monitoring [29].

## II. METHODS

### A. Fiber Fabrication

The capillary fibers were fabricated with a heat stretching process similar to that used in the fabrication of microstructured polymer optical fibers [30]. The fabrication consisted of a single step drawing process. The key to successfully using the fiber drawing method with low Young's modulus materials such as PU is a quasi-zero drawing tension [31]. Two different types of fibers were fabricated: transparent and black fibers. For the transparent fiber, the preform used was a PU tube (FB85-TPU-Clear Grayline LLC) with an outer diameter of 6.375 mm and an inner diameter of 3.175 mm. The fabrication was performed with a set drawing temperature of 240°C and a feeding velocity of 30 mm/min. The resulting capillary fibers had an outer diameter of 1.5 mm and an inner diameter of 1 mm. For the black fibers, a black PU tube (FB85-TPU-Black Grayline LLC) with an outer diameter of 6.35 mm and an inner diameter of 4.78 mm was used. The fabrication was performed with a set drawing temperature of 195°C and a feeding velocity of 40 mm/min. The resulting capillary fibers had an outer diameter of 2.5 mm, and an inner diameter of 1.7 mm. Characterization of the stress-strain response, and bend losses of the capillary are reported in the Appendix (Figs. A1 and A2).

### B. Breathing Sensor

The sensor is composed of a 20 cm long transparent PU capillary fiber, a 633 nm CW laser diode for the illumination and a Hamamatsu S5972 IR + Visible Light Si PIN photodiode to measure the optical intensity. The laser source, polymer fiber and photodetector were positioned and then fixed on two aluminum holders using commercial cyanoacrylate-based adhesive. The optical setup was then attached to a regular elastic bandage. Data acquisition was performed with an Arduino UNO board connected to a laptop.

### C. Pulse Measurement

A 3D printed frame was used to mount the detector and light source. Intermediate conventional polymer fibers couple light between the black PU sensor fiber (2 to 5 cm long) and the light source (650 nm 6 mm 5 mW laser diode) and detector (Hamamatsu S5972 IR + Visible Light Si PIN Photodiode, Through Hole TO-18). The PU fiber extends on the outside of the 3D-printed frame. The device has a screw-adjustable pad in contact with the fiber, allowing for adjustment of the pressure with which the fiber is in contact with the wrist. The power supply for the light source and the detector as well as the data acquisition were external to the wearable device. The signal was collected with a data acquisition card (National Instruments USB-4431) and processed in real time. Pulse waveforms were recorded with the sampling rate of 10 000 samples per second, which ensured that the time resolution was much higher than what was needed to determine any of the characteristic timescales of the system. The signal to noise ratio of the measurements spanned between 13 and 20 dB depending on the measurement. The system was optimized to maximize the electrical dynamic range, which was noise limited. Filtering was applied, removing frequencies below 0.2 Hz, to remove distortion due to movement, and above 45 Hz, to remove mains noise. The system was stable against temperature and power fluctuations and not affected by modal noise. The drifts were generally slow and therefore resulted in a slowly varying envelope to the signal that could be easily filtered out if necessary (see also Fig. 3A in the Appendix). The high sensitivity of the sensor and good design of the mechanics ensured that the change in amplitude was dominated by the deformation of the fiber rather than the environment.

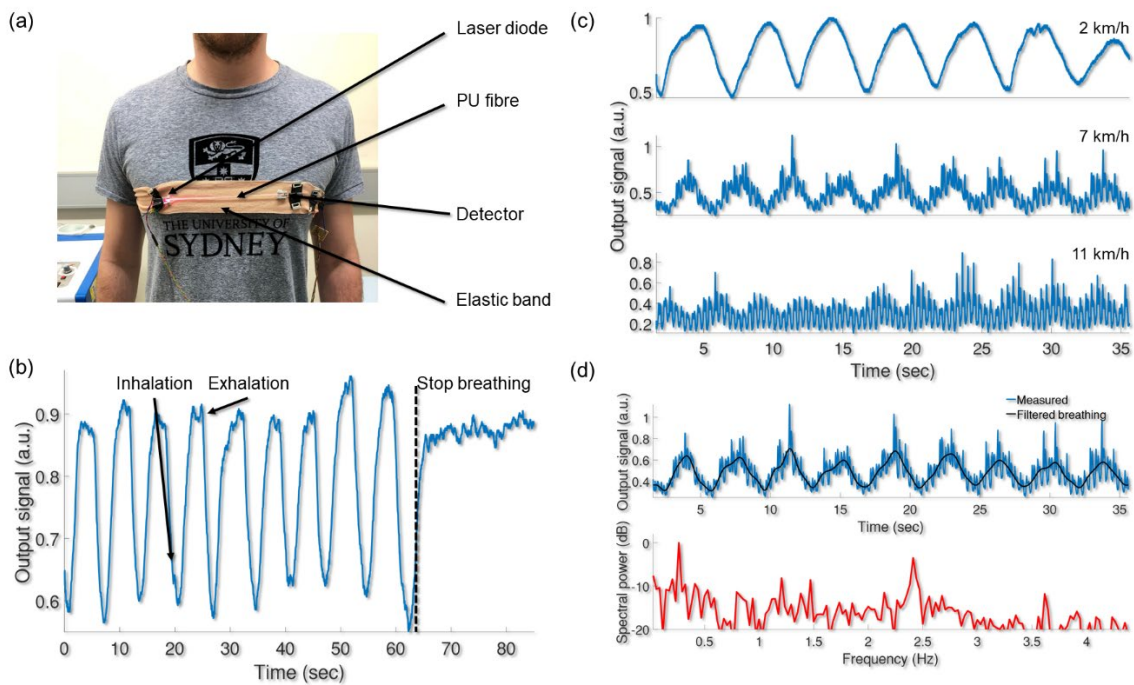

**Fig. 1.** Experimental demonstration of PU fiber wearable for measurement of respiration. (a) Photograph of the wearable on the subject; (b) measurement of respiration in resting condition; (c) measurement at various running speeds (2 km/h – top, 7 km/h – middle, 11 km/h – bottom); (d) analysis of the 7 km/h running trace.

Measurements were performed on volunteers in age groups spanning between 20 and 60 years old and repeated several times to ensure reproducibility. The testing was performed with prior approval from the University of Sydney Human Research Ethics Committee and with informed consent from all participants, as required by the relevant guidelines and regulations.

### III. RESULTS

#### A. Breathing and Foot Cadence Measured at the Chest

Respiration is a key vital parameter that can be used to monitor and improve athletic performance [32, 33] and medical monitoring. Changes in respiratory rate appear earlier compared to other vital signs such as heart rate and blood pressure [34].

The subject wore a chest strap incorporating an optical fiber sensor [Fig. 1(a)] and the intensity of the light transmitted through the fiber was monitored as a function of time. With the subject stationary, the transmission of the fiber oscillates with the periodicity of breathing, where the signal decreases during inhalation and increases during exhalation [Fig. 1(b)]. To make sure the oscillations are a true representation of the respiration, the subject held their breath and the signal transmitted through the fiber stopped oscillating.

To determine whether other factors such as movement influence the result, the sensor was tested whilst the subject was using a treadmill. Figure 1(c) shows the results for various running speed settings. The breathing periodicity can be clearly seen for all cases. As the treadmill speed increases, a higher frequency oscillation appears. Analysis of the frequency of these oscillations shows they correspond to the walking or running cadence. For instance, for 7 km/h (1.94 m/s) in Fig. 1(d) the second peak in the frequency spectrum is at 2.43 Hz, which corresponds to the subject with the step length of 0.8 m making 1.94/0.8 = 2.45 steps per second.

#### B. Pulse Measurement at the Wrist and at the Ankle

A high-fidelity measurement of the pulse over time gives access to substantial information about various health conditions. The desired information can be derived by analysis of the shape of the pulse, using the so-called pulse wave analysis (PWA) [35-44], and/or by the analysis of the statistical occurrences of the pulses, i.e., the pulse rate variability (PRV) [45-47]. With such analysis, it is possible to derive information about conditions such as hypertension [38, 39, 44], diabetes [46, 48], cardiac output [43], and mental stress [40].

This demonstration addresses the continuous monitoring of the pulse at the wrist and at the ankle. In the implementation at the wrist, the fiber is placed perpendicularly over the radial artery [Fig. 2(a)]. Light is coupled into one end of the fiber, and the changes of the optical power are monitored at the other end with a photodiode. The black PU capillary fiber used is shown in Fig. 2(b). A close-up photograph of the wearable is shown in Fig. 3.

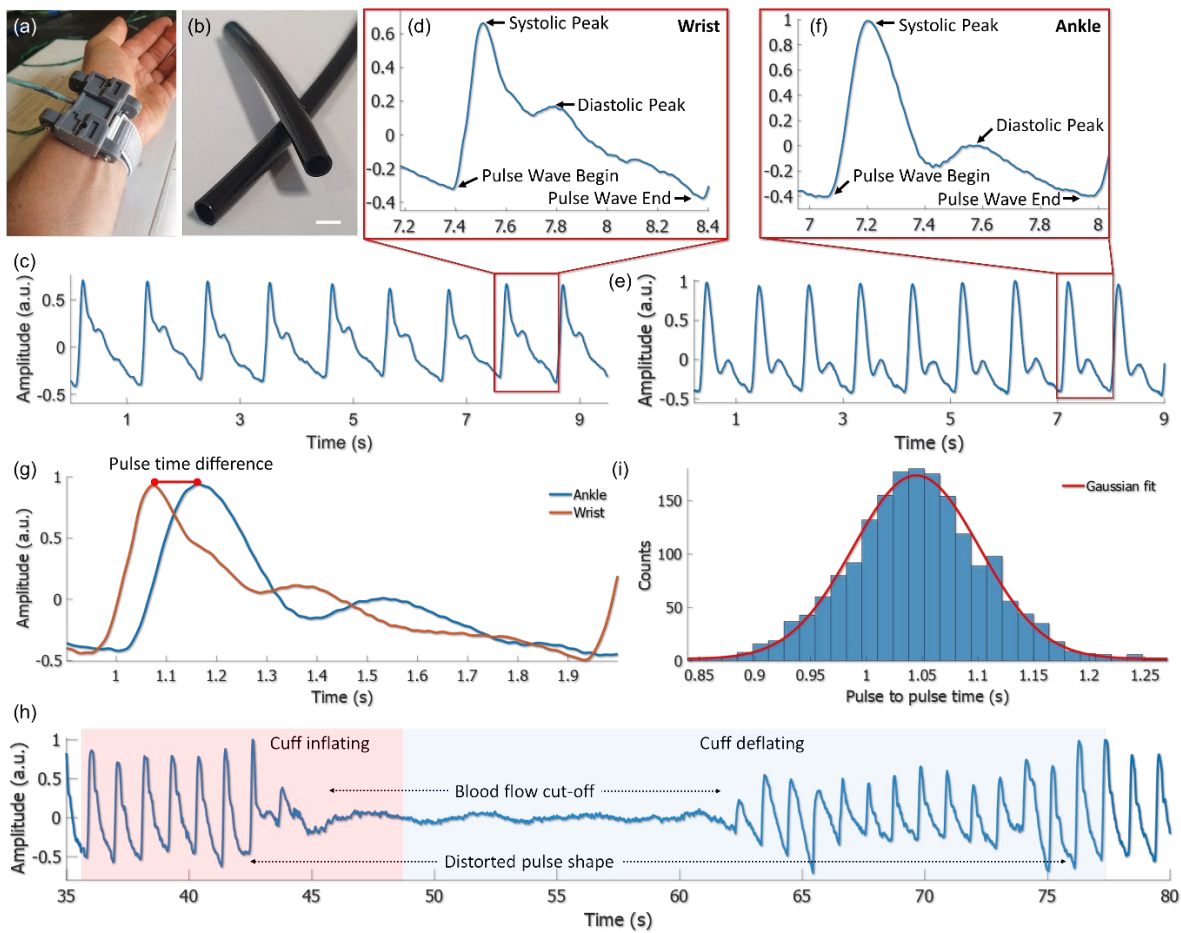

**Fig. 2.** Experimental demonstration of PU fiber wearable for measurement of the pulse. (a) Photograph of the wearable on the subject; (b) photograph of the black PU capillary fiber used – scale bar is 2 mm; (c) pulses acquired at wrist; (d) detail of a single pulse at wrist; (e) pulses acquired at ankle; (f) detail of a single pulse at ankle; (g) comparison of single pulse at wrist and ankle showing pulse time difference; (h) measured pulse during a cuff blood pressure measurement; and (i) histogram of pulse-to-pulse times over 20 minutes.

A typical recording of the pulse waveform measured by the wrist-wearable device is shown in Fig. 2(c). Detail of a single pulse is shown in Fig. 2(d). The device provides a high-resolution measurement of the pulse waveform and resolves details with clear correspondence to features of physiological significance, i.e., the foot of the pulse, the systolic and diastolic peaks of the pulse, as annotated in Fig. 2(d). The waveform allows the derivation of critical parameters of the pulse in the time domain and in the frequency domain, as well as relative amplitudes of the various features [35-53].

We performed a comparative measurement with an additional sensor applied to the ankle, simultaneously with measurement at the wrist. A typical pulse waveform at the ankle is shown in Fig. 2(e), with details of a single pulse shown in Fig. 2(f). Again, key physiological features are well resolved. The pulse shape is different in the two locations, consistent with literature [44, 49, 50], due to differences in resistance and back reflection. As we were making simultaneous measurements at the wrist and ankle it was straightforward to determine the pulse time difference, as shown in Fig. 2(g). If the two sensors were on the same artery, this time difference would be the pulse transit time. From the pulse time difference (~92 ms) and the distance between the two points on the body (~77 cm, in this case calculated as the difference between the paths from the heart to the sensors) a pulse wave velocity equivalent is readily calculated as ~8.4 m/s, which is a typical value for healthy people [49, 51].

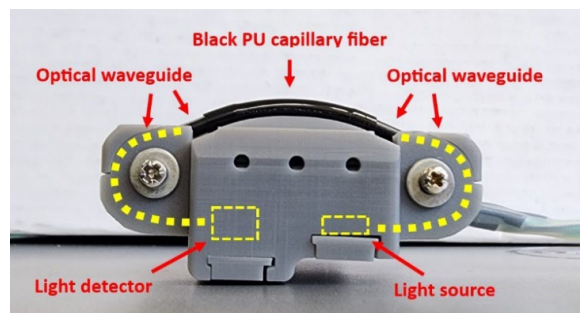

**Fig. 3.** Close-up photograph of wearable for pulse measurement. The sensitive black PU capillary fiber is coupled to the light source and detector through two polymer (PMMA) optical waveguides.

To demonstrate the sensitivity of the response to physiological changes in the actual pulse waveform, a recoding was taken while a cuff for blood pressure measurement (Omron BP5100) was inflated and deflated on the upper arm, cutting off and then releasing the blood supply to the wrist, and hence pulse signal to the sensor. The recording is shown in Fig. 2(h) and a clear deformation of the pulses in shape and amplitude is observed. Whilst our purpose here is simply to show that the sensor responds to the physiological change, analogous to stopping breathing in the previous measurement, there are some interesting features in the measured data. For example, at 42 seconds the diastolic peak disappears, while the systolic peak is still visible, before the flow of blood is cut off. Also, as the blood flow returns at 62 seconds for five beats the shape is quite different, resembling the oscillometric oscillations typical of a cuff measurement and related to the Korotkoff sounds [49, 51, 52].

We also looked at one simple, but very significant, parameter that can be obtained from the measurement: the change in time between consecutive pulses, i.e., the pulse rate variability (PRV). In the literature, PRV has been used in mental health assessment studies, in pharmaceutical research, in sleep studies, as well as in cardiovascular health and many more applications [46]. A histogram of the pulse-to-pulse times acquired in 20 minutes of recording at rest is shown in Fig. 2(i) and follows a Gaussian distribution. On average there is a new pulse every 1.05 seconds (about 57 beats per minute), with a standard deviation of 56.8 ms, which agrees with that expected for a short-term PRV in a healthy person — values of standard deviation of heart rate variability for long-term measurement, i.e., 24 h, less than 50 ms are classified as unhealthy [45].

To demonstrate that the pulse traces obtained are consistent with conventional measurements, we simultaneously measured the pulse at the wrist with the PU sensor and with a pulse oximeter (CMS-P PC Based Pulse Oximeter, Contec Medical System Co. Ltd.) on the middle finger of the same hand. The comparison is shown in Fig. 4 where the PU sensor reproduces the pulse shape with high fidelity both in its amplitude and time features, and apparently with higher resolution.

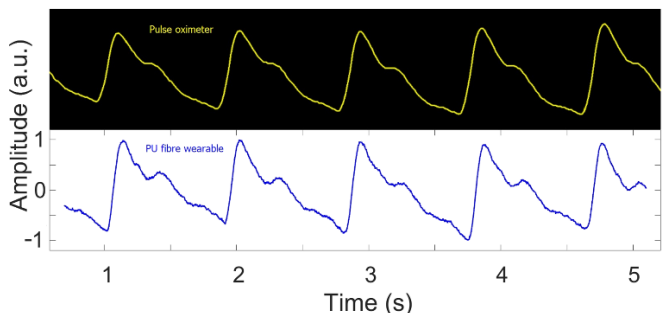

**Fig. 4.** Comparison with a commercial pulse oximeter. Waveforms measured simultaneously using pulse oximeter on the middle finger (top) and PU fiber wearable on the wrist of the same arm (bottom). The similarity of the waveforms' features indicates the physiological meaningfulness of the data obtained using the PU fiber wearable.

Measurements were performed on multiple subjects of various age groups. The results (Fig. 5) did not differ significantly in terms of the optical signal. However, physiological differences can be observed both in shape and heart rate.

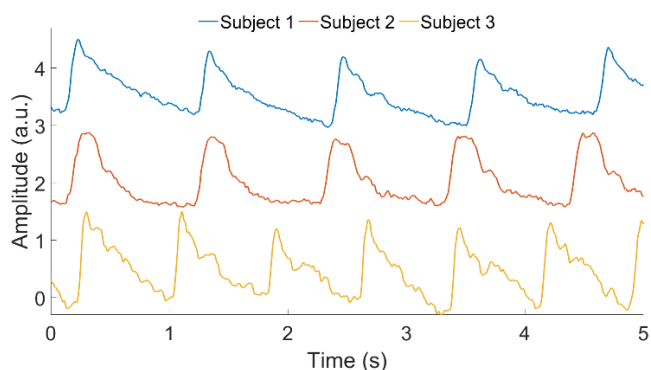

**Fig. 5.** Pulses for different age groups. Comparison of pulse waveforms acquired using PU fiber wearable at wrist for three subjects of different age groups (20-35; 35-50; 50-60 years old) showing clear differences in shape and heart rate. The traces are vertically offset for clarity.

These experiments show the immense potential of the sensing platform based on hollow-core PU fibers for cardiovascular system monitoring. The analysis of high-fidelity blood pressure waveforms obtained with the proposed fiber sensor provides a simple and inexpensive alternative to the existing heart rate monitoring technology based on a variety of much more complex optical fiber sensors [29].

## IV. DISCUSSION

The simple capillary guides used in this work are unusual. Capillaries are rarely the waveguide of choice as they are lossy, highly susceptible to perturbations, and difficult to scale to smaller dimensions as the loss increases substantially as the inner diameter reduces. However, the transmission is adequate over the length scales used in this work. If better fiber performance were desired, anti-resonant fibers would improve the transmission and allow for smaller dimensions, lower overall optical loss, and hence reduced power requirements. The sensitivity of the anti-resonant guidance mechanism to the fiber geometry suggests that these structures would also make very sensitive sensors. They have been demonstrated in soft polymers [23, 24], albeit only for THz frequencies. However, it remains a fabrication challenge to produce anti-resonant fibers in PU at small diameters, i.e., for structures operating in the visible range.

Other fiber parameters are more easily changed: the inner and outer diameters and the length of the capillary. The primary sensing mechanism in this sensor is the additional loss caused by a perturbation to the structure through external force such as bend, twist, pressure, stretch, etc. The sensitivity is thus a function of how readily the capillary is deformed, and this is determined by the Young's modulus and the inner and outer diameters — a thin-walled capillary will deform much more readily than a thick-walled capillary. Thus, the sensitivity can be engineered over several orders of magnitude by simple design changes, effectively the wall thickness. The fiber length may also be a useful design parameter, depending on whether the force is applied at a point or distributed.

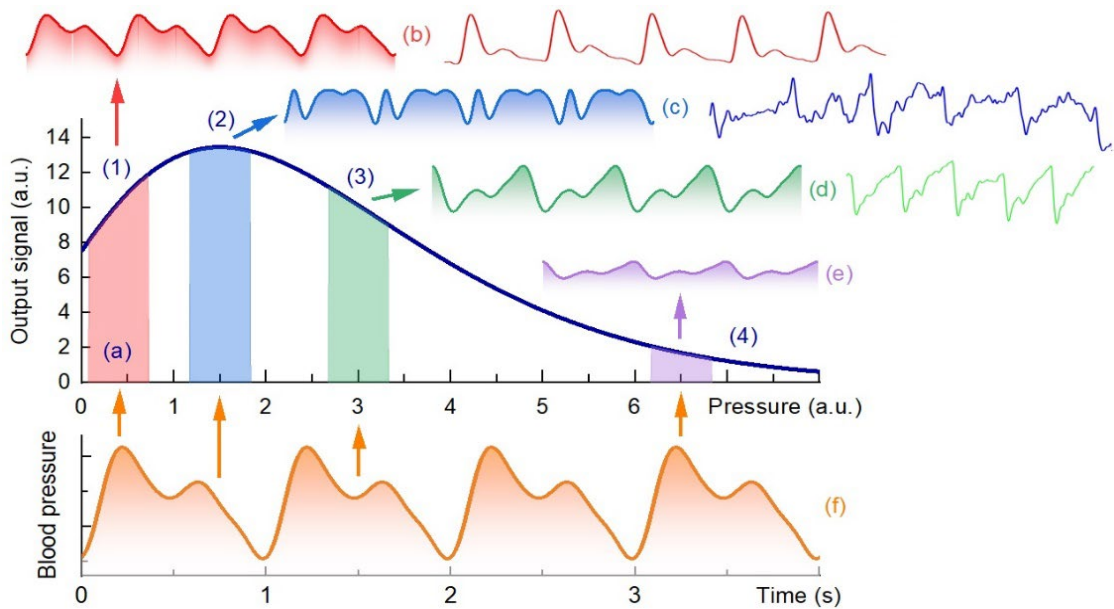

**Fig. 6.** Effect of the transfer function on the signal. (a) Transfer function of initially slightly curved TPU fiber and [(b)-(e)] effect of different parts of the transfer function on the pulse waveform (f); (b)-(d) right: measured waveforms obtained when the sensor operates in regions (1)-(3) respectively.

Another unusual aspect of the sensor is its transfer function. Intensity sensors generally show an approximately linear change in intensity over a wide scale of perturbation. However, for these fibers, depending on the sensor design, the range of approximately linear response may be limited, and indeed in some cases it may not be monotonic over the whole range. This can occur because of competing perturbations: compression and bending (both of which reduce the signal) or straightening (which increases it). If the fiber starts from an initially bent position, this can result in perturbations initially increasing the transmitted signal. Although this is another degree of freedom in the design of the sensor, in this first proof of concept care has been taken to operate in the monotonic regime.

The typical transfer function of a PU fiber sensor, with an initial small curvature such that on placing on the wrist the central portion of the sensing fiber comes into contact first, is shown in Fig. 6(a), together with the measured arterial pulse shapes associated with particular positions on the transfer function. The function has four distinct regions: the 'anomalous' quasi-linear region (1) where the output signal grows with pressure due to the partial straightening of the fiber, the intermediate highly nonlinear region (2) about the function maximum, the 'normal' quasi-linear region (3) where pressure reduces the signal due to the fiber being compressed, and the saturation region (4) with a strongly suppressed output. Only regions (1) and (3) allow accurate measurement of the arterial pulse waveform [Figs. 6(b) and 6(d)] whereas regions (2) and (4) strongly distort the waveform [Fig. 6(c)] and degrade the device sensitivity [Fig. 6(e)] respectively. Region (2) does not appear useful as the response is not monotonic. For Regions (1) to (3) actual corresponding measurements are shown on the right side of the illustration.

Optical fibers are, for obvious reasons, generally fabricated from low loss optical materials. However, the loss mechanism in capillary guidance is dominated by scattering from imperfections (including the perturbations being sensed) on the inner wall, and not by optical absorption of the material. With a capillary made of a transparent material, some fraction of that scattered light ends up being guided within the capillary wall and towards the detector, increasing the overall detected optical power, but reducing the signal to noise ratio. We found generally better performance with black PU as the capillary walls absorb any extraneous light launched or scattered into them, so that the detector is solely illuminated by light that is transmitted through the hollow core.

The sensitivity of the fibers to deformation will also result in unwanted signals that are not physiological in source. The nature of the results we have presented however allows this effect to be mitigated. All the signals we have investigated are periodic, with a characteristic periodicity and a relatively low deviation from it, i.e., narrow bandwidth. Such behavior allows us to filter out other sources of noise. The analysis of the signal in the frequency domain, connected with a large amount of collected data and aided by artificial intelligence, might also allow specific information about non-periodic features to be obtained, such as motion detection or falls.

## V. CONCLUSION

The results presented here clearly show that the novel PU fiber sensor can measure physiologically relevant signals from the body. The fidelity of the signals allows for extended and combined analysis of the data according to the different methodologies described in the literature, each of which give insight into aspects of health conditions. One of the biggest advantages of the PU fiber sensors is that they can be multiplexed, by placing multiple sensors at various positions in the body, for example allowing measurement of the pulse shape at various distances from the heart, or the simultaneous measurement of multiple physiological parameters, which can be distinguished in the frequency domain given the different characteristic frequencies. This wearable sensor network or array would allow for example, real-time relationships between activity, cardiac and respiratory performance to be established. We have shown that pulse transit time, and hence pulse wave velocity, can be obtained using this system. This opens up the possibility of a wearable capable of continuous unobtrusive monitoring of blood pressure, as well as pulse shape.

The combination of waveguide materials (low Young's modulus and color), its geometry (i.e. large air fraction), together with the sensor design (intensity based measurements, wearability and transfer function) allow the realization of unique, continuous, unobtrusive monitoring of vital signs such as breathing and opens possibilities for further applications given the large design parameter space. Moreover, integration of multiple conformable PU fibers sensors along with other wearable devices and analysis methods, such as artificial intelligence, could lead to continuous monitoring of a wide range of human physiological parameters.

## APPENDIX

### A. Stress-Strain Response of PU Capillaries

As reference for the stretchability of the material used, we report in Fig. A1 the stress-strain curve for a drawn clear PU tube with outer diameter 1.6 mm and inner diameter 1.2 mm. The measurement was performed on 50 mm long samples extended with a 20 mm/min rate on an Instron load frame. The measurement was repeated with 10 samples.

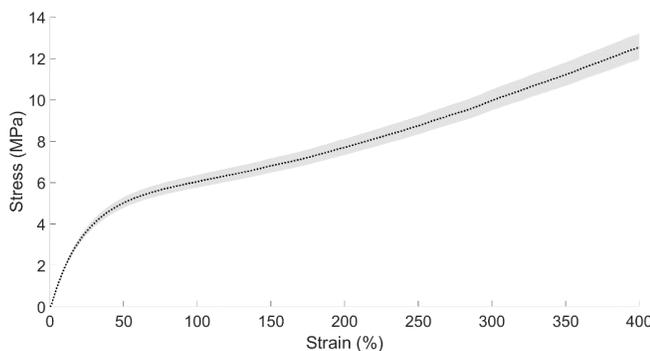

**Fig. A1.** Stress-strain response of PU capillary with 1.6 mm outer diameter and 1.2 mm inner diameter. The grey area indicates the maximum discrepancy of the measurements of the various samples.

### B. Bend Losses of Black TPU Fiber

The optical sensitivity to bending deformation can be measured as the light lost as a function of bending radius. This is one of the possible sensing mechanisms for this fiber. Figure A2 shows the attenuation vs bend radius of a 2-mm-ID, 30-mm-long black TPU fiber.

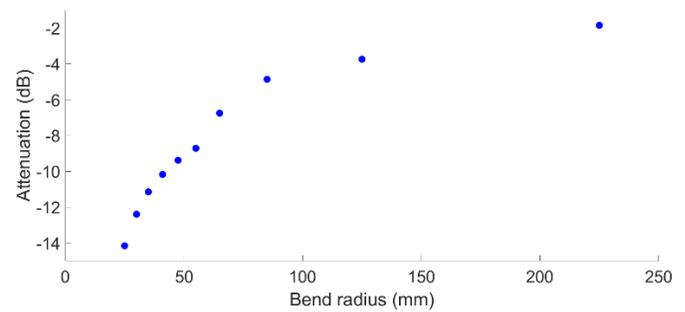

**Fig. A2.** Bend loss of black TPU fiber. Attenuation of 30-mm-long black TPU fiber versus fiber bend radius. The internal radius of the fiber is 1 mm.

### C. Stability of Sensing Data over Time

The comparison of the pulse waveforms at the beginning, in the middle, and after ten minutes of data acquisition (Fig. A3) showed no noticeable drift of the sensing data.

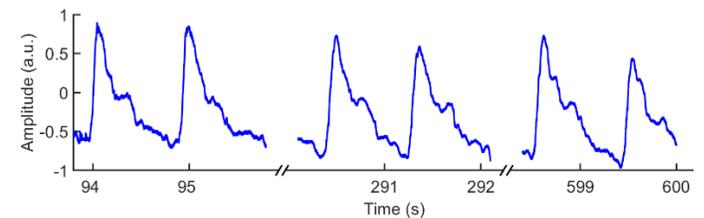

**Fig. A3.** Pulse waveforms at the beginning, in the middle, and at the end of a ten-minute measurement.

## ACKNOWLEDGMENT

Part of this work was performed at the Research & Prototype Foundry Core Research Facility at the University of Sydney, part of the Australian National Fabrication Facility (ANFF), using NCRIS and NSW State Government funding. The authors would like to acknowledge Justin Digweed for support with the fabrication, Alexander Patton for the help with the Instron measurement, Andrea Nguyen and Sean Nuttall for the help with the measurement of the transfer functions and Thomas Dunlop and Professor Clara Chow for fruitful discussions.

## REFERENCES

[1] S. Sze, D. Pan, C. R. Nevill, L. J. Gray, C. A. Martin, J. Nazareth, J. S. Minhas, P. Divall, K. Khunti, K. R. Abrams, L. B. Nellums, and M. Pareek, "Ethnicity and clinical outcomes in COVID-19: A systematic review and meta-analysis," *EClinicalMedicine*, vol. 29, 100630, 2020.
[2] G. Loke, T. Khudiyev, B. Wang, S. Fu, S. Payra, Y. Shaoul, J. Fung, I. Chatziveroglou, P.-W. Chou, I. Chinn, W. Yan, A. Gitelson-Kahn, J. Joannopoulos, and Y. Fink, "Digital electronics in fibres enable fabric-based machine-learning inference," *Nat. Commun.*, vol. 12, 3317, 2021.
[3] V. Bhatia, and A. M. Vengsarkar, "Optical fiber long-period grating sensors," *Opt. Lett.*, vol. 21, 692–694, 1996.
[4] A. D. Kersey, T. A. Berkoff, and W. W. Morey, "Multiplexed fiber Bragg grating strain-sensor system with a Fabry-Perot wavelength filter," *Opt. Lett.*, vol. 18, 1370–1372, 1993.
[5] W. Wang, N. Wu, Y. Tian, C. Niezrecki, and X. Wang, "Miniature all-silica optical fiber pressure sensor with an ultrathin uniform diaphragm," *Opt. Express*, vol. 18, 9006–9014, 2014.
[6] B. Lee, "Review of the present status of optical fiber sensors," *Opt. Fiber Technol.*, vol. 9, 57–79, 2003.
[7] J. I. Peterson and G. G. Vurek, "Fiber-optic sensors for biomedical applications," *Science*, vol. 224, 123–127, 1984.

[8]. S. Poeggel, D. Tosi, D. Duraibabu, G. Leen, D. McGrath, and E. Lewis, "Optical fibre pressure sensors in medical applications," *Sensors*, vol. 15, 17115–17148, 2015.

[9]. G. Loke, J. Alain, W. Yan, T. Khudiyev, G. Noel, R. Yuan, A. Missakian, and Y. Fink, "Computing fabrics," *Matter*, vol. 2, 786–788, 2020.

[10]. G. Loke, W. Yan, T. Khudiyev, G. Noel, and Y. Fink, "Recent progress and perspectives of thermally drawn multimaterial fiber electronics," *Adv. Mater.*, vol. 32, 1904911, 2020.

[11]. H. Bai, S. Li, J. Barreiros, Y. Tu, C. R. Pollock, and R. F. Shepherd, "Stretchable distributed fiber-optic sensors," *Science*, vol. 370, 848–852, 2020.

[12]. H. Zhao, K. O'Brien, S. Li, and R. F. Shepherd, "Optoelectronically innervated soft prosthetic hand via stretchable optical waveguides," *Sci. Robot.*, vol. 1, eaai7529, 2016.

[13]. C. Wu, X. Liu, and Y. Ying, "Soft and stretchable optical waveguide: Light delivery and manipulation at complex biointerfaces creating unique windows for on-body sensing," *ACS Sens.*, vol. 6, 1446–1460, 2021.

[14]. J. Guo, C. Yang, Q. Dai, and L. Kong, "Soft and stretchable polymeric optical waveguide-based sensors for wearable and biomedical applications," *Sensors*, vol. 19, 3771, 2019.

[15]. A. Leal-Junior, L. Avellar, A. Frizera, and C. Marques, "Smart textiles for multimodal wearable sensing using highly stretchable multiplexed optical fiber system," *Sci. Rep.*, vol. 10, 13867–13867, 2020.

[16]. J. Guo, M. Niu, and C. Yang, "Highly flexible and stretchable optical strain sensing for human motion detection," *Optica*, vol. 4, 1285–1288, 2017.

[17]. J. Guo, X. Liu, N. Jiang, A. K. Yetisen, H. Yuk, C. Yang, A. Khademhosseini, X. Zhao, and S.-H. Yun, "Highly stretchable, strain sensing hydrogel optical fibers," *Adv. Mater.*, vol. 28, 10244–10249, 2016.

[18]. Y. Qu, T. Nguyen-Dang, A. G. Page, W. Yan, T. Das Gupta, G. M. Rotaru, R. M. Rossi, V. D. Favrod, N. Bartolomei, and F. Sorin, "Superelastic multimaterial electronic and photonic fibers and devices via thermal drawing," *Adv. Mater.*, vol. 30, 1707251, 2018.

[19]. S. Shabahang, F. Clouser, F. Shabahang, and S.-H. Yun, "Single-mode, 700%-stretchable, elastic optical fibers made of thermoplastic elastomers," *Adv. Opt. Mater.*, vol. 9, 2100270, 2021.

[20]. C. K. Harnett, H. Zhao, and R. F. Shepherd, "Stretchable optical fibers: threads for strain-sensitive textiles," *Adv. Mater. Technol.*, vol. 2, 1700087, 2017.

[21]. A. Leber, B. Cholst, J. Sandt, N. Vogel, and M. Kolle, "Stretchable thermoplastic elastomer optical fibers for sensing of extreme deformations," *Adv. Funct. Mater.*, vol. 29, 1802629, 2019.

[22]. Md. R. Kaysir, A. Stefani, R. Lwin, and S. Fleming, "Measurement of weak low frequency pressure signal using stretchable polyurethane fiber sensor for application in wearables," in *Conference on Electrical Information and Communication Technology*, 2017.

[23]. A. Stefani, S. C. Fleming, and B. T. Kuhlmey, "Terahertz orbital angular momentum modes with flexible twisted hollow-core antiresonant fiber," *APL Photonics*, vol. 3, 051708, 2018.

[24]. A. Stefani, J. H. Skelton, and A. Tuniz, "Bend losses in flexible polyurethane antiresonant terahertz waveguides," *Opt. Express*, vol. 29, 28692–28703, 2021.

[25]. E. A. J. Marcatili and R. A. Schmeltzer, "Hollow metallic and dielectric waveguides for long distance optical transmission and lasers," *Bell Syst. Tech. J.*, vol. 43, 1783–1809 1964.

[26]. R. F. Cregan, B. J. Mangan, J. C. Knight, T. A. Birks, P. ST. J. Russell, P. J. Roberts, and D. C. Allan, "Single-mode photonic band gap guidance of light in air," *Science*, vol. 285, 1537–1539, 1999.

[27]. W. Belardi and J. C. Knight, "Hollow antiresonant fibers with reduced attenuation," *Opt. Lett.*, vol. 39, 1853–1856, 2014.

[28]. J. C. Travers, T. F. Grigorova, C. Brahms, F. Belli, "High-energy pulse self-compression and ultraviolet generation through soliton dynamics in hollow capillary fibres," *Nat. Photonics*, vol. 13, 547–554, 2019.

[29]. R. He, L. Shen, Z. Wang, G. Wang, H. Qu, X. Hu, R. Min, "Optical fiber sensors for heart rate monitoring: A review of mechanisms and applications," *Results in Optics*, 2023, https://doi.org/10.1016/j.rio.2023.100386

[30]. M. C. J. Large, L. Poladian, G. Barton, and M. A. van Eijkelenborg, (eds) "Microstructured Polymer Optical Fibres," Springer, 2008.

[31]. R. Lwin, S. Fleming, A. Stefani, and Md. R. Kaysir, "Fiber forming process," Australian patent 201932993, 2019.

[32]. A. Nicolò, C. Massaroni, and L. Passfield, "Respiratory frequency during exercise: The neglected physiological measure," *Front. Physiol.*, vol. 8, 922, 2017.

[33]. M. J. Gross, D. A. Shearer, J. D. Bringer, R. Hall, C. J. Cook, and L. P. Kilduff, "Abbreviated resonant frequency training to augment heart rate variability and enhance on-demand emotional regulation in elite sport support staff," *Appl. Psychophysiol. Biofeedback*, vol. 41, 263–274, 2016.

[34]. A. Dix, "Respiratory rate: the benefits of continuous monitoring," *Nursing Times* [online], vol. 114, 36–37, 2018.

[35]. J. N. Cohn, "Noninvasive pulse wave analysis for the early detection of vascular disease," *Hypertension*, vol. 26, 503–508, 1995.

[36]. M. F. O'Rourke, A. Pauca, and X.-J. Jiang, "Pulse wave analysis," *Br. J. Clin. Pharmacol.*, vol. 51, 507–522, 2001.

[37]. Z. Fan, G. Zhang, and S. Liao, in Advanced Biomedical Engineering, eds. G. D. Gargiulo and A. McEwan, Ch. 2 *Pulse Wave Analysis*, IntecOpen, 2011.

[38]. S.-H. Liu, L.-J. Liu, K.-L. Pan, W. Chen, and T.-H. Tan, "Using the characteristics of pulse waveform to enhance the accuracy of blood pressure measurement by a multi-dimension regression model," *Appl. Sci.*, vol. 9, 2922, 2019.

[39]. American Heart Association News, "Pulse waves may detect problems missed by blood pressure readings," 2019, https://www.heart.org/en/news/2019/07/29/pulse-waves-may-detect-problems-missed-by-blood-pressure-readings.

[40]. P. H. Charlton, P. Celka, B. Farukh, P. Chowienczyk, and J. Alastruey, "Assessing mental stress from the photoplethysmogram: A numerical study," *Physiol. Meas.*, vol. 39, 054001, 2018.

[41]. J. P. Mynard, A. Kondiboyina, R. Kowalski, M. M. H. Cheung, and J. J. Smolich, "Measurement, analysis and interpretation of pressure/flow waves in blood vessels," *Front. Physiol.*, vol. 11, 1085, 2020.

[42]. P. J. Aston, M. I. Christie, Y. H. Huang, and M. Nandi, "Beyond HRV: attractor reconstruction using the entire cardiovascular waveform data for novel feature extraction," *Physiol. Meas.*, vol. 39, 024001, 2018.

[43]. B. Saugel, K. Kouz, T. W. L. Scheeren, G. Greiwe, P. Hoppe, S. Romagnoli, and D. de Backer, "Cardiac output estimation using pulse wave analysis—physiology, algorithms, and technologies: A narrative review," *Br. J. Anaesth.*, vol. 126, 67–76, 2021.

[44]. X. Sun, F. Su, X. Chen, Q. Peng, X. Luo, and X. Hao, "Doppler ultrasound and photoplethysmographic assessment for identifying pregnancy-induced hypertension," *Exp. Ther. Med.*, vol. 19, 1955–1960, 2020.

[45]. F. Shaffer, and J. P. Ginsberg, "An overview of heart rate variability metrics and norms," *Front. Public Health*, vol. 5, 258, 2017.

[46]. E. Mejía-Mejía, J. M. May, R. Torres, and P. Kyriacou, "A. Pulse rate variability in cardiovascular health: a review on its applications and relationship with heart rate variability," *Physiol. Meas.*, vol. 41, 07TR01, 2020.

[47]. E. Mejía-Mejía, J. M. May, M. Elgendi, and P. A. Kyriacou, "Differential effects of the blood pressure state on pulse rate variability and heart rate variability in critically ill patients," *npj Digit. Med.*, vol. 4, 82, 2021.

[48]. K.-M. Liao, C.-W. Chang, S.-H. Wang, Y.-T. Chang, Y.-C. Chen, and G.-C. Wang, "Assessment of cardiovascular risk in type 2 diabetes patients by insight into radial pulse wave harmonic index," *Acta Physiol.*, vol. 228, 2019.

[49]. J. M. Padilla, E. J. Berjano, J. Saiz, L. Facila, P. Diaz, and S. Merce, "Assessment of relationships between blood pressure, pulse wave velocity and digital volume pulse," *Comput. Cardiol.*, vol. 2006, 893–896, 2006.

[50]. L. A. Geddes, *Handbook of Blood Pressure Measurement*. Springer Science & Business Media, New York, 2013.

[51]. A. Díaz, C. Galli, M. Tringler, A. Ramírez, and E. I. Cabrera Fischer, "Reference values of pulse wave velocity in healthy people from an urban and rural Argentinean population," *Int. J. Hypertens.*, vol. 2014, 653239, 2014.

[52]. M. Shimizu, "Blood flow in a branchial artery compressed externally by a pneumatic cuff," *J. Biomech. Eng.*, vol. 114, 78–83, 1992.

[53]. B. Wilk, R. Hanus, J. Novosad, and P. Dančová, "Blood flow in the brachial artery compressed by a cuff," *EPJ Web of Conferences*, vol. 213, 2099, 2019.

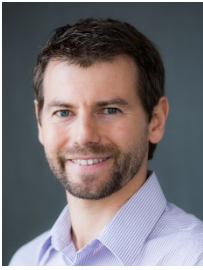

**Alessio Stefani** received a B.Sc. and M.Sc. degrees in physics engineering from the Politecnico di Milano, Italy in 2006 and 2008, and a PhD in Photonics Engineering from the Technical University of Denmark, Denmark in 2012.

In 2012 he was a Postdoc at the FEMTO-ST Institute, Université de Franche-Comté, France. From 2012 to 2014 he was a Research Fellow at the Max Planck Institute for the Science of Light, Germany. From 2014 to 2015 he was a Postdoc at the Technical University of Denmark, Denmark. From 2015 to 2016 he was Postdoc at the University of Sydney, Australia. From 2016 to 2019 he was a Marie Curie Fellow at the Technical University of Denmark, Denmark and at the University of Sydney, Australia. Since 2019 he is at the Fraunhofer Institute for Integrated Circuits IIS, Germany, where he covers the role of Chief Scientist and he is a Research Affiliate at the University of Sydney, Australia. He is an author of more than 30 journal articles, 70 conference papers and 2 patents. His research interests are in specialty optical fibers, as well as metamaterials and integrated optoelectronics circuits.

Dr. Stefani is a Senior Member of Optica and a member of the Australian and New Zealand Optical Society.

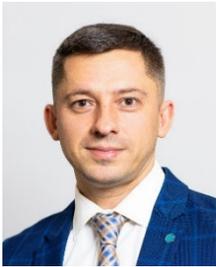

**Ivan D. Rukhlenko** received B.Sc. and M.Sc. degrees in physics from Peter the Great Saint Petersburg Polytechnic University, Russia in 2001 and 2003, and PhD and Habilitation in optics from ITMO University, Russia in 2006 and 2016.

From 2006 to 2008 he was Lecturer with ITMO University, Saint Petersburg, Russia. From 2008 to 2015 he was a Research Fellow and DECRA Fellow with the Electrical and Computer Systems Engineering Department, Monash University, Victoria, Australia. From 2015 to 2019 he was Head of Modelling and Design of Nanostructures Laboratory at ITMO University, Saint Petersburg, Russia. From 2019 he is Research Fellow in Photonic Structures and Devices at the University of Sydney, Sydney, New South Wales, Australia. He is the author of more than 160 journal articles, three books, two book chapters, and one patent. His research interests include biomedical applications of optical fibers, metamaterials, and optics of quantum nanostructures.

**Antoine F. J. Runge** received a B.Sc. and M.Sc. degrees in Physics from the Université de Franche-Comté, France in 2009 and 2011, respectively. He then obtained his PhD in Physics from the University of Auckland in 2015.

From 2015 to 2017 he was a Research Fellow within the Optoelectronics Research Centre at the University of Southampton, UK. In 2018 he joined the University of Sydney, Australia, where he is currently a Senior Research Fellow. In 2022, he was awarded an Australian Research Council DECRA Fellowship. His research focuses on optical solitons in mode-locked lasers and microresonators.

**Maryanne C. J Large** has a B.Sc. from the University of Sydney (1990) and PhD from Trinity College Dublin (1994).

Following her PhD she did post-doctoral fellowships at l'Ecole Supèrieure de Physique et de Chimie Industrielles de la Ville de Paris and the Commissariat à l'Energie Atomique (as a Marie Curie Fellow). From 1997-2000 she was a lecturer in Physics at the Dublin Institute of Technology, before returning to Australia to join the Optical Fibre Technology Centre, and then the School of Physics at the University of Sydney, where she led the group that developed the first polymer microstructured fibres. From 2011- 2013 she was a Research Manager at Canon Information Systems Research Australia, returning to the University of Sydney in 2013. She is the author of 1 book, edited one book, 3 book chapters, 80 refereed journal papers, and 77 refereed conference papers, 5 patents.

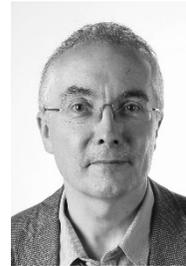

**Simon C. Fleming** received a B.Sc. degree in electrical and electronic engineering from The University of Leeds, UK in 1982, and a PhD in electrical and electronic engineering from The University of Leeds, UK in 1987.

From 1982 to 1983 he was a Design Engineer with Racal SES, Tewkesbury, England. From 1986 to 1991 he was a Research Fellow in The Department of Electrical and Electronic Engineering, The University of Leeds, UK. From 1991 to 1993 he was a Senior Research Fellow at BTLabs, Ipswich, UK. From 1994 he has held several roles at The University of Sydney, Sydney, New South Wales, Australia, including Director of the Optical Fibre Technology Centre (1999-2008), Professor of Physics (2009-) and Academic Director of the Research and Prototype Foundry Core Research Facility (2017-). He is the author of 130 journal articles, 230 conference papers, one book chapter and nineteen patents. His research interests are primarily specialty optical fibres, from design and fabrication to application, most recently to biomedical devices.

Prof. Fleming is a Fellow of the The Institution of Engineering & Technology, a Senior Member of Optica, a member of the Australian and New Zealand Optical Society, and a Chartered Engineer.